\documentstyle[preprint,aps]{revtex}
\def\lsim{\mathrel{\rlap{\lower4pt\hbox{\hskip1pt$\sim$}}
    \raise1pt\hbox{$<$}}}         
\def\gsim{\mathrel{\rlap{\lower4pt\hbox{\hskip1pt$\sim$}}
    \raise1pt\hbox{$>$}}}         
\begin{document}
\draft
\preprint{ADP-93-216/T134}
\title{Pion form factors
at intermediate momentum transfer \\ in a covariant approach}
\author{Adam Szczepaniak\cite{byline1}}
\address{Department of Physics,
Florida State University, Tallahassee, FL 32306\\
and\\
Department of Physics, 
North Carolina State University, Box 8202, 
Raleigh NC 27695}
\author{Anthony G. Williams\cite{byline2}}
\address{Department of Physics and SCRI, Florida State University,
Tallahassee, FL 32306\\
and\\
Department of Physics and Mathematical Physics,
University of Adelaide, SA 5005, Australia}
\maketitle
\begin{abstract}
We study the pion electromagnetic and $\gamma^* + \pi^0 \to \gamma$
transition form factors at intermediate momentum transfer.
We calculate soft, nonperturbative
corrections to the leading perturbative amplitudes which arise
from the $q\bar q$-component of the pion wave function. We
work in Minkowski space and use a Lorentz covariant, gauge-invariant
generalized perturbative integral representation for the $q\bar q$
amplitudes.
For the transition form factor we find relative
insensitivity to the detailed
nonperturbative structure of the wavefunction for $|q^2|\gsim 10$~GeV$^2$,
whereas considerable sensitivity is found for the electromagnetic form
factor.
\end{abstract}
%
\pacs{12.38.Aw, 12.38.Lg, 12.40-y}

%
\section{Introduction}
\label{intro}
There has recently been considerable attention focused on the 
applicability (or otherwise) of the leading twist perturbative
QCD (pQCD) analysis in 
descriptions of exclusive processes at large momentum transfers.
Theoretical studies of such processes are typically justified on
the basis of the factorization
of the hard, perturbative and the soft, nonperturbative contributions to
hadronic matrix elements.
This factorization
appears explicitly through the operator product expansion
(OPE)~\cite{OPE}, where amplitudes are determined by
the diagonal or almost diagonal matrix elements of nonlocal operators.
It is applied to far-off-diagonal (i.e., large momentum transfer)
matrix elements of local operators in a parton
model description of exclusive hadronic form
factors~\cite{LEBR,EFRA,CHZH1,CHZH2}.
The largest discrepancy between experimental data and the leading
order perturbative calculations is observed in the
predictions for absolute normalization of hadronic contributions to
cross  sections such as the K factor in the Drell-Yan process and
electromagnetic form factors. This may be due to large
corrections from
next to leading order (in $\alpha_s$) terms and/or contributions
from operators with higher dimension (twist).  For example,
the pion
electromagnetic (e.m.) form factor [$F(Q^2)$]
at large spacelike momentum transfer has been the subject
of several studies, which showed that there were
difficulties associated with the application of leading order pQCD
arising from the appearance
of intermediate states with small virtuality.
These arise from the
end-point  region in the loop integration over the quark
momenta~\cite{CHZH2,ISLS,DIRA}
in what are necessarily assumed to be hard amplitudes.
In particular, they may lead to large
perturbative corrections, rendering the perturbative expansion
inconsistent. Recent analyses~\cite{SZMA,STLI} have shown that
the inclusion of Sudakov-type corrections and transverse momenta in the
analysis can suppress the unwanted end-point contributions and lead to a self
consistent perturbative expansion for momentum transfers as low as a few GeV.
There is, however, an open question on the role of the higher twist
operators. This point was first emphasized in Ref.~\cite{ISLS},
and to the best of our knowledge for the case of the pion e.m. form factor
has only been
examined in a model independent way in Refs~\cite{GETE} and ~\cite{IOSM}.
In Ref.~\cite{GETE} the contributions from twist-3 operators in a parton model
factorization approach has been calculated and shown to be even larger
than the na\"{\i}ve asymptotic one for $Q^2 \lsim 20$ GeV$^2$. In
Ref~\cite{IOSM} the behaviour of the form factor at $Q^2 \sim 10$
GeV$^2$ has been calculated within the QCD sum rule approach~\cite{SHVAZA}
and was also shown to be dominated by contributions from higher dimension
operators.
In this work we wish to address the question of the role of
nonperturbative corrections to descriptions based both on parton model
factorization and the OPE. 
In Sec.~\ref{pionff-sec} we use a Lorentz covariant
wavefunction formalism to extend the parton model approach to the
pion electromagnetic form factor.
This is an extension of
initial work presented elsewhere\ \cite{SZWI}.
In Sec.~\ref{g-pi-g-sec} we calculate
the nonperturbative corrections  to the $\gamma^* \pi \to \gamma$ form
factor  [$T(Q^2)$] as an example of an application of the OPE techniques.
We give our conclusions in Sec.~\ref{concl}.
\section{Pion electromagnetic form factor}
\label{pionff-sec}
A complete representation of the pion e.m. form factor, $F(Q^2)$
is given in Fig.~\ref{pionampl-fig}
and involves a pointlike coupling of the photon
to a quark and anti-quark, (all dressings etc. are contained in $M^\mu$); 
$M^\mu=F(Q^2)(P^\mu+ P'^\mu)$, represents the matrix element of the e.m.
quark current,
$J^\mu(0) = \; :{\overline \psi}(0)e_q\gamma^\mu\psi(0):$, between pion states 
of momenta $P$ and $P+q\equiv P'$, ($Q^2\equiv-q^2$). In QCD, as the momentum
transfer, $Q^2\to \infty$, the leading contribution to the form factor
factorizes into a convolution of a hard scattering kernel $(T)$ which
involves a minimal number of hard, highly virtual parton lines,
with soft distribution amplitudes ($\phi$) for the initial and final pion.
Assuming factorization, the $q{\overline q}$ state contribution
to the form factor in $M^\mu$  can be written as
\cite{LEBR,EFRA}, \begin{equation}
F(Q^2) \to F(Q^2)_{\rm pQCD} = \int_{-1}^{1} dxdy
{\overline \phi}_A(x,\lambda^2)
T(x,y;Q^2,\lambda^2)\phi_A(y,\lambda^2)\;, 
\label{fact}
\end{equation}
where $\lambda$ refers to the factorization scale and where the subscript
$A$ on $\phi_A$ indicates the soft pion amplitude
arising from an axial-vector spinor structure.  In pQCD,   
large and small virtuality contributions from loops are factorized into
$T$ and $\phi$ respectively.
$x$ and $y$ are fractions of momentum relative to the 
momentum of the initial and final on-shell massless pion
respectively. The hard scattering amplitude
to leading order in $\alpha_s$ is given by
a Born diagram with a one gluon exchange projected on a collinear,
massless, on-shell $q{\overline q}$ state with an axial-vector spinor
structure in initial and final channels (i.e., the twist-2 contribution).
Similarly, the corresponding distribution amplitudes $\phi_A$ are obtained
by a collinear projection of the axial-vector ($\Phi_A$) component of the
matrix element,
\begin{equation}
\Phi_{\alpha\beta}(p,P)
\equiv \int d^4z\; e^{i(p-{1\over 2}P)
\cdot z}\langle 0|T\psi^i_\alpha(0)
P^{ji}(z,0) \bar\psi^j_\beta(z) |P\rangle\;. \label{Phi-PT}
\end{equation}
In Eq.~(\ref{Phi-PT}) sums over the color indices $i$ and $j$ are
understood and the parallel transport operator can be written as
\begin{equation}
P(x+z,x) \equiv {\cal P}\exp[ig_s\int_0^1 ds\; z\cdot A(x+sz)]\;,\label{PT}
\end{equation}
where ${\cal P}$ denotes the path-ordering operator and $\alpha$ and $\beta$
are spinor indices.
The color labels $i$ and $j$ in Eq.~(\ref{Phi-PT}) have the ordering shown
since the parallel-transport operator usually appears in the combination
$\bar\psi(y)P(y,x)\psi(x)$, where the matrix summation over
color indices is implied.
The gluon field operators are denoted by $A_a^\mu(x)$, where
$a=1,2,\dots,8$ and where $A^\mu(x) \equiv (1/2)\lambda_a A_a^\mu(x)$.
$g_s$ is the strong coupling constant.
 In the amplitude $\Phi$ the incoming pion has 4-momentum
$P$, the quark has 4-momentum $p+(P/2)$, and the antiquark has
4-momentum $p-(P/2)$.
We see that the parallel transport operator plays the same role in $\Phi$
as that played by the gluon flux tube in the valence quark model
in that it ensures local color gauge invariance~\cite{string}.
The color-singlet amplitude $\Phi$ is identical to the familiar Bethe-Salpeter
(BS) amplitude  when the $q$ and $\bar q$ couple at a single
point, (i.e., when $z=0$).  This is just the situation that occurs, for
example, in a calculation
of the pion decay constant $f_\pi$, where there is a coupling
to a pointlike $W$ or $Z$, (see later).

We shall estimate the size of  ${\cal O}(1/Q^2)$ corrections
to the asymptotic behaviour $F_{\rm pQCD}$  of the form factor
once the collinear projection has been removed; these corrections are
due to the presence of nonzero quark virtualities in the soft
$\Phi_A$ part of the  wave function in  Eq.~(\ref{Phi-PT}), where
\begin{equation}
\Phi(p,P) \equiv \gamma_5 \Phi_P(p,P) 
+ {\rlap\slash \!P}\gamma_5 \Phi_A(p,P)
+ {\rlap\slash \!p}\gamma_5 \Phi_{A'}(p,P)
+ [\,{\rlap\slash \!p},{\rlap\slash \!P}]\gamma_5\Phi_T(p,P)\;.
\label{Phi}
\end{equation}
Here $\Phi_i(p,P)$ for $i=P$, $A$, $A'$, and $T$ are scalar functions
of $p^2,(p\cdot P)$, and $P^2$.  The subscripts $P$, $A$, $A'$, and $T$
denote the spinor matrix structure of the particular component
of $\Phi$.  The three functions $\Phi_{P}$, $\Phi_{A}$,
and $\Phi_{T}$
are even functions of $p\cdot P$, whereas $\Phi_{A'}$ must be an odd
function of $p\cdot P$.  We will work here in the chiral limit where
the pion is massless and so for an on-shell pion we
have $P^2=M^2_\pi=0$.

A standard assumption~\cite{NAK} is that the scalar functions $\Phi_i(p,P)$
can be written in terms of a generalized perturbation-theory integral
representation, which for equal mass particles we can express as
\begin{eqnarray}
\Phi_i(p,P)&=& \sum_{\beta=2}^{\infty}\int_{-1}^{1}
dx\int_0^\infty d\nu^2 \nonumber\\
&&\times{{\bar g_{\beta i}(x,\nu;p,P)}
\over { \left[ \left(p - x {1\over 2}P\right)^2 
- \left[\nu^2 +m^2- {1\over 4}(1-x^2)P^2 \right]
+ i\epsilon\right]^\beta} },
\label{Phi-pert}
\end{eqnarray}
where $m$ is the $u$ and $d$ quark mass
(assumed degenerate for simplicity) and the ${\bar g}$ are 
scalar functions
generally taken to be polynomially dependent on $p^\mu$~\cite{NAK}.
In the above $\beta$ is an integer $\geq 2$.
Once the perturbative corrections are taken into account, the renormalized
parameters
and the scalar functions ${\bar g}$ pick up a  $\log\lambda^2$ dependence,
where the factorization scale 
$\lambda$ divides soft and hard contributions. We shall return
to this point later. In order to ensure confinement in such a representation
we can require, for example, that the functions $\bar g$
have no support [i.e., $\bar g(x,\nu,p,P)=0$] in the region
where $\nu^2\leq (P^2/4)-m^2$.
In the chiral limit for on-shell pions ($m=0$)
the surviving nonperturbative scale is that
typical of dynamical chiral symmetry breaking and confinement.
We will denote this scale here  by $\mu$ and we clearly expect
$\mu\lsim1$~GeV,  at $\lambda \sim 1$~GeV.

With this motivation in mind we now make an assumption
about the form of the functions $\bar g_{\beta}$, i.e., we assume
\begin{equation}
\bar g_{\beta i}(x,\nu;p,P)=
\delta(\nu^2-\mu^2)\;N_{\beta i}\;g_{\beta i}(x)
\;,
\label{gbar}
\end{equation}
where the mass scale $\mu$,
and the functions $g_{\beta i}(x)$
have yet to be specified.
The functions $g_{\beta i}$ will be chosen to be dimensionless and so 
the normalizations $N_{\beta i}$ are included for dimensional reasons, 
and are to be determined by fitting the pion
wave function to some physical quantities, (e.g., $f_\pi$ as discussed
later). With this choice we then find from Eq.~(\ref{Phi-pert}) that
\begin{equation}
\Phi_i(p,P) = \sum^\infty_{\beta=2} N_{\beta i} \int_{-1}^{1} dx 
{g_{\beta i}(x)
\over 
[(p - x{1\over 2}P)^2-\mu^2+i\epsilon]^\beta}\;.
\label{Phi-model}
\end{equation}
As already pointed out, only the soft part of the amplitudes $\Phi_i(p,P)$ are
to be used in the
calculation for the form factors and
has to contribute only at small virtualities, $(p\pm {1\over 2}P)^2
< \lambda^2$, 
with the contributions from hard virtualities  being factorized into the
hard scattering.  In order for that to happen, we require that
the soft part of the amplitude $\Phi_i$ given by Eq.~(\ref{Phi-model})
has its short distance
behaviour softer than the one corresponding to a high momentum
gluon exchange which leads to the
hard scattering amplitude.
In other words, any large relative momentum behaviour in 
Eq.~(\ref{Phi-model}) that would be identical to
that from a hard gluon exchange 
has to be subtracted, in order to avoid double counting. 
 
In order to isolate that part of $\Phi(p,P)$ which
corresponds to the general 
light-cone $z^2 < 1/\lambda^2$ short distance behaviour of the matrix
element in Eq.~(\ref{Phi-PT}),
one has to study the $p^2_\perp > \lambda^2$ behaviour of the
corresponding light cone projection
$\Phi^{LC}(p^+,p_\perp,P)$ of $\Phi_A$,
since the dominant asymptotic behavior arises from this.
We use standard notation,
where $z^\pm\equiv z^0\pm z^3$, $z_\perp\equiv
(z^1,z^2)$,  etc., and work
in the infinite momentum frame where $P\to(P^+,0^-,0_\perp)$.
As usual, we also define the fraction of relative longitudinal
momentum as $x\equiv 2p^+/P^+$, [we will see shortly that this $x$ can be
related to the $x$ in Eq.~(\ref{Phi-model})].
The light cone wave function is defined as~\cite{LEBR}
\begin{equation}
\Phi^{\rm LC}(p^+,p_\perp,P) =
\int {{dp^-}\over {2\pi i}}\; 
 \Phi_A(p,P)\;.
\label{Phi-LC}
\end{equation}
Substituting the expression given by Eq.~(\ref{Phi-model}) for $\Phi_A$ in 
Eq.~(\ref{Phi-LC}) we obtain for $p^2_\perp > \lambda^2$,
\begin{equation}
\Phi^{\rm LC}(p^+,p_\perp,P) \propto
\sum_{\beta=2}
{{N_{\beta A}g_{\beta A}(x)}
\over {P^+\left(p^2_\perp+\mu^2\right)^{\beta-1}}}
 = {{N_{2 A} g_{2 A}(x)}\over {p_\perp^2P^+}}
\left[\delta_{\beta,2} +
{\cal O}\left({{\mu^2}\over {\lambda^2}}{{\lambda^2}\over
{p_\perp^2}}\right)\right]. \label{Phi-model-LC}
\end{equation}
On the other hand the pQCD expansion of $\Phi^{\rm LC}$ for
$p^2_\perp > \lambda^2$ corresponding to a hard gluon exchange is given
by~\cite{BFLS}
\begin{eqnarray} 
&&\Phi^{\rm LC}_{\rm pQCD}(x,p_\perp) \propto
{{\alpha_s(p_\perp^2)}\over {p^2_\perp P^+}}
\sum_{n} a^{(\lambda)}_{n} \left( {{\log(p^2_\perp/\Lambda^2_{QCD}) }
\over {\log(\lambda^2/\Lambda^2_{QCD}) }}
\right)^{-(\gamma_{n}-2\gamma_F)/2\beta_0} (1-x^2) G^{3/2}_{n}(x)
\nonumber\\ &&a^{(\lambda)}_{n} =
P^+\int_{-1}^{1}dx'\int_{0}^{\lambda^2}{{dp'^2_\perp}
\over{16\pi^2}}\int{{dp'^-}\over {2\pi i}}
 \Phi_A(x',p'_\perp,p'^-)
G^{3/2}_{n}(x') \nonumber \\
&&\gamma_{n} = 2C_F
\left(1 + 4 \sum_{j=2}^{n+1} {1\over j} 
- {2\over{(n+1)(n+2)}} \right)\, , \,\gamma_F = C_F\;,
\label{Phi-pQCD}
\end{eqnarray}
where $C_F=4/3$ and $\beta_0=11 - {2\over 3}n_f$ with $n_f$ being the
number
of quark flavors. The coefficients $a^{(\lambda)}_{n}$ contain all of the
nonperturbative
information and are sensitive to the behaviour of the soft LC wave
function i.e. for $p_\perp^2 < \lambda^2$.
The functions $G_n^{3/2}(x)$ are the usual Geigenbauer polynomials
of order $3/2$ and contain all of the longitudinal momentum information.

An examination of the double integral (over $x$ and $p^-$) 
encountered in deriving Eq.~(\ref{Phi-model-LC}) shows that
the $p^-$-integral will give zero unless we have $x=2p^+/P^+$.
Hence as previously stated we can
identify the $x$ appearing in the functions
$g_{\beta i}(x)$
with the fraction of longitudinal momenta.
Comparing the leading terms in the
$\lambda^2/p^2_\perp$ expansion of Eqs.~(\ref{Phi-model-LC}) and
(\ref{Phi-pQCD}),
we see that leading power behaviour of the $\beta=2$
term in Eq.~(\ref{Phi-model-LC}) is the same as the one given by a hard gluon
exchange tail in Eq.~(\ref{Phi-pQCD}). Thus the $\beta=2$
term should be excluded from the soft pion wave function
in this limit, as explained above. Furthermore, comparing the $x$-dependence
of the corresponding asymptotic terms, we see that 
if  logarithmic corrections of Eq.~(\ref{Phi-pQCD})
were taken into account in
Eqs.~(\ref{Phi-model}) and (\ref{Phi-model-LC}), one would have
for $p^2_\perp > \lambda^2$,
\begin{equation}
g_{2 A}(x) \to g_{2 A}(x,p^2_\perp) \\
\propto (1-x^2)\sum_{n} a^{(\lambda)}_{n} \left(
{{\log(p^2_\perp/\Lambda^2_{QCD}) }
\over {\log(\lambda^2_\perp/\Lambda^2_{QCD}) }}
                           \right)^{-\gamma_{n}/2\beta_0} G^{3/2}_{n}(x)\; ,
\label{evol}
\end{equation}
On substituting Eq.~(\ref{Phi-model}) into Eq.~(\ref{Phi-pQCD}) we
also see that the coefficients
$a_n^{(\lambda)}$'s can be expressed in turn in terms of the
(normalized) Geigenbauer moments of the
functions $g_{\beta A}(x)$, for which
$g_{\beta A}(x)\to g_{\beta A}(x,p^2_\perp)$ for $p_\perp^2 \lsim
\lambda^2$.  This relationship can then be inverted using the orthonormality
of the Geigenbauer polynomials [with the measure $(1-x^2)$] to give
the $g_{\beta A}$ in terms of the coefficients $a_n^{(\lambda)}$.
The $\beta=2$ component in the evaluation of the coefficients
$a_n^{(\lambda)}$
(for $p_\perp^2 < \lambda^2$) must be excluded in Eq.~(\ref{Phi-pQCD})
because it would generate a nonperturbative
model-dependent cutoff, [i.e., a factorization scale
($\lambda$) dependence], for physical amplitudes like $f_\pi$ for example.
Thus, in keeping with the usual concept of factorization we have for
$\beta\ge 3$
\begin{equation}
g_{\beta A}(x)\to g_{\beta A}(x,\lambda^2) \propto
(1-x^2) \sum_n a_n^{(\lambda)} G^{3/2}_n(x). 
\label{g3}
\end{equation}
Hence, from
Eqs.~(\ref{Phi-model-LC}) and (\ref{evol}) it follows that
$g_{2 A}(x,p_\perp^2)$ should be identified with
the pQCD evolved quark distribution amplitude.
For the nonperturbative ($\beta\ge 3$) components 
of the wave function ($\beta\ge 3$) we
will assume here that such terms  can be adequately
represented by the piece that will dominate the 
$\lambda^2/p_\perp^2$ expansion, i.e., we neglect
$\beta\ge 4$ and use a $\beta=3$ piece alone.
This is not essential, but
obviously simplifies the analysis without altering the 
asymptotic behaviour. We shall comment further on this point later.

The functions $g_{3 A}(x,\lambda^2)$ can then be assumed to
correspond to typical soft quark distribution amplitudes. 
Following the above arguments we will simplify
the notation. We shall now
understand that $\Phi_i(p,P)$ of Eq.~(\ref{Phi-model})
refer to {\it only}
the soft nonleading contributions with $\beta=3$.
Hence we write in place
of Eq.~(\ref{Phi-model}) for $\Phi_i$ 
\begin{equation}
\Phi_i(p,P)_{(\lambda)} =  N_{3 i}\int_{-1}^{1} dx
{g_{3 i}(x,\lambda^2)
\over 
[(p - x{1\over 2}P)^2-\mu^2+i\epsilon]^3}\;.
\label{Phi_A}
\end{equation}
The distribution amplitudes are typically normalized such that
\begin{equation}
\int_{-1}^1dx \;g_{\beta i}(x,\lambda^2)=1
\label{g-norms}, 
\end{equation}
and in particular for $p_\perp^2/\lambda^2 \to \infty$
the usual {\it asymptotic} result for $g_{2 A}$ is
\begin{equation}
g_{2 A}(x) \to g_{2 A}(x,\infty) = {3\over 4}(1-x^2)\;. 
\label{g-infty}
\end{equation}
while the soft quark distribution amplitude might, for example, be taken to be
parameterized
in terms of the Chernyak and Zhitnitsky moments~\cite{CHZH2}
\begin{equation}
g_{3 A}(x,\lambda^2) \to g_{3 A}(x,\sim 1\mbox{ GeV}^2)\simeq
g_{\rm CZ}(x)\equiv {15\over 4}(1-x^2)x^2 \;.
\label{g_CZ} \end{equation}
The normalization constant $N_{3 A}$ can be fixed by the
fact that the 0-th moment of $\Phi$ is 
directly related to 
the pion decay constant $f_\pi=93$ MeV. 
Only $\Phi_A$ and $\Phi_{A'}$ can contribute to $f_\pi$ and we shall for
convenience make a further simplifying assumption, which is that
nonperturbative $\Phi_{A'}$ piece can be neglected. 
A straightforward calculation of $f_\pi$ then only involves $\Phi_A$ and gives
\begin{equation}
f_\pi= -{3\over 8\pi^2\mu}
\int_{-1}^1dx[N_{3 A}g_{3 A}(x,\lambda^2)
]\;.
\label{f-pi}
\end{equation}
It follows from Eq.~(\ref{g-norms}) that $N_{3 A}=
-(8\pi^2/3)f_\pi\mu^2$.
This leads to our final expression for $\Phi_A$,
\begin{equation}
\Phi_A(p,P)_{(\lambda)}
= -
{{8\pi^2f_\pi}\over 3}
\int_{-1}^{1} dx {{g_{3 A}(x,\lambda^2)}\mu^2
\over { \left[ \left(p - x {1\over 2}P\right)^2 
- \mu^2 + i\epsilon\right]^3} }\;.
\label{Phi-vec}
\end{equation}
It is worth noting that Eqs.~(\ref{Phi-model-LC}), (\ref{Phi-pQCD}),
and (\ref{g-norms}) guarantee  
that $f_{\pi}$ is factorization scale (i.e., $\lambda$) independent
even when the pQCD corrections are included.
If we were to generalize to a wave
function containing $\beta > 3$ terms, then the unknown 
structure functions, $g_{\beta i}(x,\lambda^2)$, ($\beta=4,\cdots$) and
corresponding scales, $\mu_{\beta i}$,
would need to be fixed from the normalization of additional physical matrix
elements. Matrix elements involving 
higher number of covariant derivatives such as
\begin{equation}
\langle 0|{\overline \psi}(0)(i\stackrel{\leftarrow}{D})^2\gamma^\mu\gamma_5
(i\stackrel{\leftrightarrow}{D})^{\mu_1}
\cdots (i\stackrel{\leftrightarrow}{D})^{\mu_n}\psi(0) | P
\rangle_{(\lambda)}
= if_\pi \langle k_\perp^2 \rangle^A  \langle x^n \rangle_{G}
P^\mu \cdots P^{\mu_n}
\end{equation}
involve the moments of the distribution amplitude
$g_G(x,\lambda^2)$~\cite{CHZH2},
\begin{eqnarray}
& & g_G(x,\lambda^2)P^\mu = \int d(z\cdot P) e^{-x(z\cdot P)}{5\over 9}
\langle 0| {\overline \psi}({z\over 2})g_s
\tilde G_{\mu\nu}(0)\gamma^\nu\gamma_5 \psi(-{z\over 2}) | P
\rangle_{(\lambda)}, \nonumber \\
& & \langle x^n \rangle_G \equiv \int dx x^n g_G(x,\lambda^2)\;.
\end{eqnarray}
These are associated with the quark longitudinal 
momentum distribution in the presence of an additional gluon operator
$\tilde{G}_{\mu\nu}= {1\over
2}\epsilon_{\mu\nu\alpha\beta}G_{\alpha\beta}$.
As described
in the introduction, however, we restrict our analysis to
the contribution
from the $q{\overline q}$ sector only and thus are ignoring possible
higher Fock sector distribution amplitudes.
 
Returning now to the e.m. form factor calculation, the standard quark counting
rules 
and the pQCD formula of Eq.~(\ref{fact}) to leading order in $Q^2$ 
predict $F(Q^2)\to F_{\rm pQCD}(Q^2) \sim 1/Q^2$ as $Q^2\to\infty$. 
The $1/Q^2$ behavior
comes from diagrams with a single gluon exchange in a quark loop. In order
to maintain QCD gauge invariance, once gluon degrees of freedom 
have been explicitly introduced, one should include the soft 
spectator gluon diagram of Fig.~\ref{oge-fig}b) along with the
hard gluon exchange diagrams of Fig.~\ref{oge-fig}a).
However, only Fig.~\ref{oge-fig}a) 
contributes to the leading $1/Q^2$ behaviour and so the gauge dependence 
enters as an ${\cal O}(1/Q^2)$ correction to the leading behaviour. 
In order to show that the formalism presented here gives the right 
asymptotic behaviour we calculate the diagrams of Fig.~\ref{oge-fig}a)
using the $\Phi_A$ contribution and an arbitrary covariant gauge where
the gluon propagator has the form
$D_{\alpha\beta} =
[-g_{\alpha\beta} + (1-a) k_\alpha
k_\beta/(k^2+i\epsilon)]/(k^2+i\epsilon)$.
We shall refer to this contribution to the form factor as $F_A$.
This gives then ($q^2\equiv -Q^2\leq 0$ and $P'\equiv P+q$)
\begin{eqnarray}
& & i(P + P')^\mu F_A(Q^2) =
-6\int{d^4\ell\over (2\pi)^4}{d^4k\over (2\pi)^4} \nonumber \\ 
& & \mbox{tr}\Bigg[
\Big({2\over 3}\Big) {\overline \Phi}_A\Big(k+{P'\over 2},-P'\Big)
T^\mu(k,l,P,P')\Phi_A\Big(\ell+{P\over 2},P\Big)S^{-1}_F(\ell)\nonumber\\
&+&
\Big({-1\over 3}\Big) \Phi_A\Big(\ell-{P\over 2},P\Big)T^\mu(k,l,-P,-P')
{\overline\Phi_A}\Big(k-{P'\over 2},-P'\Big)S^{-1}_F(\ell)
\Bigg],
\label{FF-eq}
\end{eqnarray}
where $(2/3)$ and $(-1/3)$ are the $u$ and $d$ quark charges
respectively and where for the pion\cite{GOSOGU,LS} 
\begin{equation}
{\overline\Phi}(p,P) = T\Phi^T(p,-P)T^{-1} = \Phi(p,P)
\;.\label{Phi-bar}
\end{equation}
The transpose acts on the spinor indices and $T=i\gamma^1\gamma^3=
T^\dagger=T^{-1}$ is the time-reversal
operator in the usual Dirac representation.
The hard scattering amplitude $T^\mu$ describing the piece of the diagrams
which do not involve the wave functions is given by,
\begin{eqnarray}
T^\mu(k,l,P,P')  & = & g_s^2[\gamma^\alpha S_F(k+P)\gamma^\mu] \otimes 
[\gamma^\beta]D_{\alpha\beta}(k-l) \nonumber \\
& + &
[\gamma^\mu S_F(l+P')\gamma^\alpha]\otimes
[\gamma^\beta]D_{\alpha\beta}(k-l). \label{sc}
\end{eqnarray}
With $S_F$ being the perturbative quark propagator. 
Using the form for $\Phi_A$ given by Eq.~(\ref{Phi-vec})  one finally
obtains, %
\begin{eqnarray} 
& & F_A(Q^2) = {{2 g_s^2 f^2_\pi}\over {9}}
\int_{-1}^{1} dx g_{3A}(x,\lambda^2) \int_{-1}^{1} dx' g_{3A}(x',\lambda^2)
\nonumber \\ & & \Bigg[\int_{0}^{1} d\xi {{\mu^2 Q^2 \xi^3 (1-x)}\over
{\left[\xi(1-x){{Q^2}\over 2} + \mu^2\right]
\left[\xi(1-\xi)(1-x)(1-x'){{Q^2}\over 4} + \mu^2\right]^2}}
\nonumber \\ &+& (x \to x')\Bigg]
\left[1 + {\cal O}\left((1-a){{\mu^2}\over {Q^2}}\right)
\right]\;. \label{F_A}
\end{eqnarray}
Once all of the one loop corrections
to the Born diagram are taken into account and renormalization
at $\mu^2_R \sim Q^2$ is performed [in order to cancel
potentially large logs of the form $\log({Q^2/\mu_R^2})$], the
unrenormalized coupling, $g_s^2$ gets replaced by $4\pi\alpha_s(Q^2)$.
Similarly, logs of the form $\log({Q^2/\lambda^2}$), with $\lambda$
introduced in a loop calculation in order to factorize soft and hard
contributions, are responsible for the evolution of the initial, soft
distribution amplitudes $g_{3A}(x,\lambda^2)$ in a form that is given by
Eq.~(\ref{evol})~\cite{CHZH2}.
Thus the expression for the renormalized form factor $F_A$ 
can be obtained from Eq.~(\ref{F_A}) by the following
replacement,
\begin{equation}
g^2_s g_{3A}(x,\lambda^2) g_{3A}(x',\lambda^2) \to 4\pi \alpha_s(Q^2)
g_{2A}(x,Q^2) g_{2A}(x',Q^2) \left[1 + O(\alpha_s)\right]\;.
\label{corr}
\end{equation}
The remaining,
finite perturbative corrections in Eq.~(\ref{corr}) are actually
large~\cite{DIRA}, however
we are concerned only with the role of the nonperturbative corrections
and so in the following we do not discuss the perturbative ones.
The ${\cal O}(\mu^2/Q^2)$ terms on the RHS come from the
gauge fixing terms proportional to $(1-a)$
and we can neglect these contributions here.
In the asymptotic limit the 
leading contribution from the integral in Eq.~(\ref{F_A}) to the
renormalized form factor is of the form 
$F_A(Q^2) \to F_{\rm pQCD}(Q^2)$, where
\begin{equation}
F_{\rm pQCD}(Q^2) = 
{{64\pi\alpha_s(Q^2) f^2_\pi}\over {9Q^2}}
\left[\int_{-1}^{1} dx {{g_{2A}(x,Q^2)}\over {1-x}}\right]^2
\;.\label{F-pQCD}
\end{equation}
When we also substitute the asymptotic form for $g_{2 A}$ from
Eq.~(\ref{g-infty}) we recover the well known
leading-twist pQCD result~\cite{LEBR,EFRA,CHZH1}.
As mentioned in the Introduction the twist-3 operators were shown to give 
also a sizable contribution to the e.m. form factor at intermediate momentum 
transfers. These corrections correspond to the contributions from the 
$\Phi_P$ and $\Phi_T$ pieces of the our wavefunction and since they mix 
under renormalization~\cite{GETE} both have to be included.
We use the following generalization of the collinear twist-3 wave
function introduced in Ref.~\cite{GETE} that contains {\it both}
$P$ and $T$ components,
\begin{equation}
\Phi_{PT}(p,P) = N_{3P}\int_{-1}^{1} dx {{ g_{3P}(x,\lambda^2) } \over
{ [(p -x{1\over 2}P)^2 - \mu^2 + i\epsilon]^4 } }\Big[(-{1\over 2}
{\rlap \slash\!P}
+{\rlap \slash \!p}) \gamma_5 ({1\over 2}{\rlap \slash \!P} +
{\rlap \slash\!p})\Big] \;.
\end{equation}
We write this as a $\beta=3$ piece because the extra power
of momentum squared in the numerator cancels a corresponding  
power in the denominator.
We use $\int_{-1}^{1} dx g_{3P}(x) = 1$ and the normalization coefficient 
$N_{3P}$ calculated from the matrix element
\begin{equation}
\langle 0| {\overline \psi}(0)\gamma_5 \psi(0) | P \rangle_{(\lambda)}
= i \sqrt{2} { { \langle {\overline q} q \rangle }_{(\lambda)} \over
{f_\pi}},
\end{equation}
where  $\langle {\overline q} q \rangle _{\lambda\sim
1\mbox{GeV}} \sim -(250 \mbox{MeV})^3$ is the quark condensate.
The contribution $F_P(Q^2)$ to the form factor coming from the twist-3
wavefunction above can be obtained in a way analogous to the $F_A(Q^2)$
one and is given by,
\begin{eqnarray}
& & F_P(Q^2) = { { 8\pi \alpha_s(Q^2)
 \langle {\overline q}q \rangle^2_{(Q)} } \over
{ 9f_\pi^2} }
\int_{-1}^{1} dx g_{2P}(x,Q^2) \int_{-1}^{1} dx'
g_{2P}(x',Q^2)S(x,x')       x
\int_{0}^{1}   d\xi \int_{0}^{1}d\eta \eta \nonumber \\
& &\times\left[ \left(1 - \xi { {1-x} \over 2} - (1 - \xi) { {1-x'}\over 2}
\right) - 1 \right] \Biggl[ { 1\over  { [ (1-\xi) { {1-x'}\over 2} Q^2
(1-\eta + \xi\eta  { {1-z} \over 2} ) + \mu^2]^3 } }
+ (x \leftrightarrow x') \Biggr] \nonumber \\
\label{F_P}
\end{eqnarray}
with~\cite{GETE,SHVAZA}
\begin{equation}
\langle {\overline q}q \rangle_{(Q)} =
\left( { {\alpha_s(Q^2)} \over {\alpha_s(\lambda^2)} }
\right)^{-{4/\beta_0}}
\langle {\overline q}q \rangle_{(\lambda)}
\end{equation}
and $g_{2P}(x,Q^2)$ being the twist-3 distribution amplitude
obtained from the $g_{3P}(x,\lambda^2)$ one
through the pQCD evolution equation coming from the renormalization of
the vacuum-to-pion matrix element of the twist-3
operator~\cite{GETE}
\begin{eqnarray}
& & g_{2P}(x,p^2_\perp)
\propto \sum_{n} b^{(\lambda)}_{n} \left(
{{\log(p^2_\perp/\Lambda^2_{QCD}) }
\over {\log(\lambda^2_\perp/\Lambda^2_{QCD}) }}
\right)^{-\tilde{\gamma}_{n}/2\beta_0}
P_{n}(x), \nonumber \\
& & \tilde{\gamma}_{n} = 2C_F
\left(1 + 4 \sum_{j=2}^{n+1} {1\over j}
\right), \nonumber \\
& & g_{3P}(x,\lambda^2) = \sum_n b^{(\lambda)}_n P_n(x),
\label{evolp}
\end{eqnarray}
and $P_{n}$ being the Legendre polynomials.
Finally
\begin{equation} S( x x' ) =  \left( {{\sigma^2} \over {Q^2} }
 \right)^{ { {C_F}\over {\beta_0}}
\log { {\alpha_s(\sigma^2) } \over {\alpha_s(Q^2)} } }
\;, \sigma^2 = x x' Q^2 + \mu^2
\end{equation}
is the Sudakov form factor~\cite{CHZH2,GETE}.
Although we have so far been neglecting the $O(\alpha_s)$ perturbative 
corrections, the Sudakov form factor, $S(\sigma^2/Q^2)$ is
needed in Eq.~(\ref{F_P}) to keep $Q^4 F_P(Q^2)$ finite in the $Q^2 \to 
\infty$ limit as required by perturbation theory ~\cite{GETE}.
 
In our treatment the pion e.m. form factor is
$F(Q^2)=F_A(Q^2)+F_P(Q^2)$,
where $F_A$ is given by by Eqs.~(\ref{F_A}) and (\ref{corr})
and $F_P$ is given by Eq.~(\ref{F_P}).
We have used a running coupling $\alpha_s =
4\pi/\beta_0\log(Q^2/\Lambda_{QCD})$  with $\Lambda_{QCD}=150\mbox{MeV}$.
For the pseudoscalar distribution 
amplitude an asymptotic form $g_{2P}(x,Q^2) \to g_P(x,\infty) = 1/2$ has
been used.
For $g_{2A}$ we have used the Chernyak and
Zhitnitsky moments of Eq.~(\ref{g_CZ}), i.e.,
$g_{2A}(x,Q^2)\simeq g_{3A}(\lambda^2)
\simeq g_{\rm CZ}(x)$ on the basis of Eq.~(\ref{corr}) and
since the $Q^2$ dependence due to the pQCD evolution is small.
This is certainly reasonable since the finite $O(\alpha_s)$ corrections to
the Born amplitude have been neglected as well ~\cite{DIRA}.
In our calculation the parameter $\mu$
has been set to be 
$\mu=338\mbox{MeV}$ which is a 
mass-scale typical of both confinement
and dynamical chiral symmetry breaking and it has
been chosen such that our pion wavefunction give a correct normalization
of the $\gamma^* \pi \to \gamma$ form factor which we analyze
in the next section. 
In Fig.~\ref{pionff-fig} we have shown four curves, where
the lower solid curve is the prediction resulting from the above arguments.
For comparison we also show the result for
$F_{\rm pQCD}$ defined in Eq.~(\ref{F-pQCD}) 
when we use $g_{2A}(x,Q^2)=g_{CZ}(x)$,
which is the upper solid curve. 
The dashed curves correspond to the solid curves, except that we
have used the asymptotic distribution $g_{2A}(x,Q^2)\to g_{2A}(x,\infty)$
given in Eq.~(\ref{g-infty})
rather than $g_{\rm CZ}$.  The dashed curve which
increases at low $Q^2$ is the one which contains $F_P$.
At very large $Q^2$ all results converge as is of course
to be expected.
With the asymptotic form for $g_{2A}$ neither the pQCD formula nor the form
factor calculated here appear to reproduce the data. In the Chernyak
and Zhitnitsky case, which is more realistic for a distribution amplitude
at $\lambda \sim 1\mbox{GeV}$, the introduction of the soft corrections in
the wavefunction
significantly reduce the size of the form factor below the pure 
pQCD result in the intermediate regime ($\simeq 5$ to 20~GeV$^2$).
In Fig.~\ref{rff} we illustrate the overall $Q^2$ dependence of the 
{\cal O}(1/$Q^4)$ (at large $Q^2$) corrections to the pQCD form
factor $F_{pQCD}$.  We attempt to do this by plotting the dimensionless
combination,
\begin{equation}
\delta F(Q^2) \equiv { { F_{\rm pQCD}(Q^2) - F_A(Q^2) }\over 
{F_{\rm pQCD}(Q^2) } } + F_P(Q^2)\;.
\end{equation}
The net effect is a substantial correction at $Q^2 \sim
10\mbox{GeV}^2$ for both the Chernyak and Zhitnitsky distribution amplitude
(solid line) and the asymptotic one (dashed line). 

\section{$\gamma^* \pi^0\to\gamma$ form factor}
\label{g-pi-g-sec}
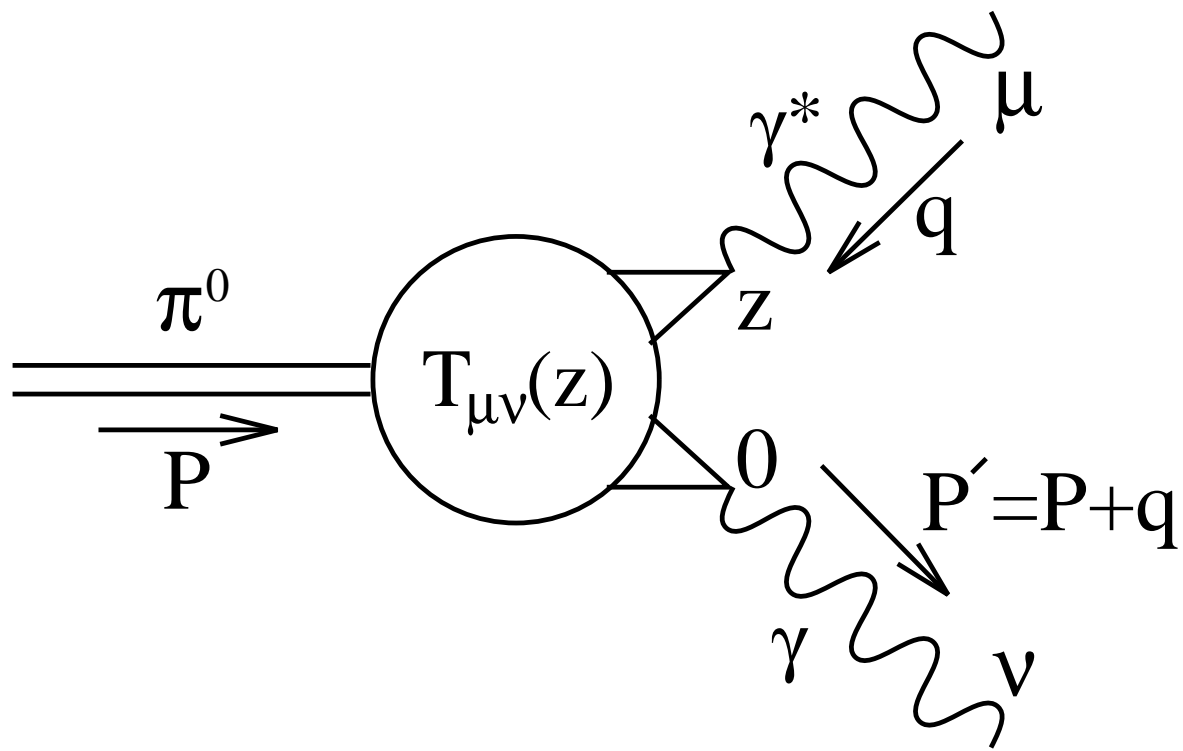
\begin{figure}
\caption{Amplitude for the $\gamma^*\pi \to \gamma$ from factor}
\label{gpgfig}
\end{figure}

The full representation of the $\gamma^*\pi\to\gamma$ form factor amplitude
$T_{\mu\nu}(Q^2)$ where $Q^2 \equiv -q^2$ and $q^2$ is the mass of the
virtual photon is shown in Fig.~\ref{gpgfig}.
The relevant hadronic piece of the corresponding amplitude is given
by~\cite{DACH}
\begin{equation}
T_{\mu\nu} = i\int d^4z e^{-iq\cdot z}\langle 0|TJ_\mu(z)J_\nu(0)|P
\rangle = \epsilon_{\mu\nu\rho\sigma}q^\rho P^\sigma T(Q^2)\;,
\label{T-mu-nu}
\end{equation}
%


For $|Q^2|\to\infty$ an operator product expansion (OPE) of the two
currents in terms 
of local, gauge invariant operators can be used to calculate $F(Q^2)$. 
We wish to extrapolate the calculation of 
the form factor from the asymptotic region, where only the twist-2
piece of the OPE of the product $J_\mu(z)J_\nu(0)$ and the light-cone wave 
function of Eq.~(\ref{Phi-pQCD}) contribute~\cite{DACH}, down to 
the intermediate region of momentum transfer. Furthermore, we wish to
do this while maintaining the valence ($q \bar q$) contribution
only.  This requirement
corresponds to keeping particular contributions to the matrix
element $\langle0|TJ_\mu(z)J_\nu(0)|P\rangle$.
This includes contributions of the form
$S\left(\partial_{\mu_0}\dots\partial_{\mu_m}{\overline \psi}
\gamma^{\mu_{m+1}}\gamma_5D_{\mu_{m+2}}\dots D_{\mu_{m+n}}\psi - 
\mbox{traces}\right)$,
(which are twist-2 pieces that lead to the known perturbative 
result~\cite{DACH}),  contributions from the twist-2 ``trace'' pieces that 
correspond to replacing one or more pairs of covariant
derivatives $D^\mu D^\nu$ by $g^{\mu\nu}D^2$, and finally
higher twist contributions 
obtained by contracting pairs of covariant derivatives without 
the spin-2 $g^{\mu\nu}$ tensors. 
For example,
$S\left(\partial_{\mu_0}\dots\partial_{\mu_m}{\overline \psi}
\gamma^{\mu_{m+1}}\gamma_5D_{\mu_{m+2}}\dots D^2\dots
D_{\mu_{m+n}}\psi \right)$ is a twist-4 contribution. 
$S$ represents the usual symmetrization of the space-time indices.
Clearly, there are no explicit gluons or sea-quarks in these operators.
The sum of all such operators is equivalent to the
replacement~\cite{SHVA,MU}
\begin{equation}
\langle 0|TJ_\mu(z)J_\nu(0)|P\rangle \to
2 \epsilon_{\mu\nu\sigma\rho}[\partial^\sigma \Delta(z)]
\mbox{Tr}\left[\gamma^\rho\gamma_5
\langle 0|T{\overline \psi}(z)P(z,0)\psi(0)|P\rangle^{(\lambda)}\right]
\label{JJ}
\end{equation}
where $P(z,0)$ is the parallel transport operator and
$\Delta(z)\equiv 1/(4\pi^2 z^2-i\epsilon)$ originates from
the vector part of a free massless Dirac propagator. 
A simple description of this is that we have a purely valence
pion wavefunction with a photon attached to each of the $q$ and $\bar q$
legs and with a massless quark line connecting the two photon vertices.
This is shown in Fig.~\ref{gpgpqcd}.
The replacement represented by Eq.~(\ref{JJ}) corresponds to assuming
that all nonleading contributions can be
subsumed into the valence pion wavefunction, which we immediately identify
with the $\Phi_A$ component of the covariant wavefunction from
Sec.\ref{pionff-sec}. Thus we shall drop the index $A$ on the
wavefunction. The leading (hard) term dominates asymptotically
and as in the case of the pion e.m. form factor [see Eq.~(\ref{fact})] is
obtained through a collinear projection of the wave function in
Eq.~(\ref{JJ}). From Eqs.~(\ref{Phi-pQCD}), (\ref{evol}), (\ref{T-mu-nu}), and
(\ref{JJ}) the pQCD behaviour is given by  $T(Q^2)\to T_{\rm pQCD}(Q^2)$,
where 
\begin{equation}
T_{\rm pQCD}(Q^2) = {{4f_\pi}\over {3Q^2}}\int_{-1}^{1} dx
 {{g_2(x,Q^2)}\over {1 - x}} \;. \label{TpQCD}
\end{equation}
In general, for our soft wave function of Eq.~(\ref{Phi-vec}), we find that
Eqs.~(\ref{T-mu-nu}) and (~\ref{JJ}) lead to
\begin{equation}
T(Q^2) = {{4f_\pi}\over {3Q^2}}\int_{-1}^{1} dx
{{g_3(x,\lambda)}\over {1-x}}
\left[1 - {{2\mu^2}\log\left({1 +{Q^2\over 2\mu^2}(1-x) 
- i\epsilon}\right)
\over {Q^2(1-x)}}\right]\;,
\label{FF}
\end{equation}
which is the form that we use below the factorization scale
($|Q^2|<\lambda^2$) with $g_3$ some soft quark distibution amplitude.
Above the factorization scale ($|Q^2|>\lambda^2$) the perturbative
corrections to the c-number coeficient [$\Delta(z)$]
in Eq.~(\ref{FF}) lead to the replacement
\begin{equation}
g_3(x,\lambda^2) \to g_2(x,Q^2)
\end{equation}
through Eq.~(\ref{evol}) with the coefficients $a^{(\lambda)}_n$ given by
Eq.~(\ref{g3}).  Thus, above the factorization scale we simply use the
same distribution amplitude as below, but include
the logarithmic correction factor of Eq.~(\ref{evol}).

In Figs.~\ref{g-pi-g-fig}a) and b) we plot our predictions for the
$T(Q^2)$ form factor in the spacelike region ($Q^2>0$) and timelike
region ($Q^2<0$) respectively. In the spacelike (timelike) regions the upper
(lower) solid line is the pQCD prediction of Eq.~(\ref{TpQCD}) with the
Chernyak and Zhitnitsky distribution amplitude used as an input for the
soft distribution amplitude, i.e., 
($g_2(x,Q^2)\to g_{\rm CZ}(x)$) for $|Q^2| \leq
\lambda^2 = 1\mbox{GeV}^2$. For $|Q^2| > \lambda^2$ pQCD as explained above
the evolution given by
Eq.~(\ref{evol}) was used to generate perturbative logarithmic corrections
to $T(Q^2)$. The upper (lower) dashed curves represent the pQCD predictions
in the spacelike (timelike) regions for the asymptotic distribution
amplitude [see Eq.~(\ref{g-infty})] (for which pQCD evolution has been
neglected).  Our prediction for the form factor [see Eq.~(\ref{FF})]
using the Chernyak and Zhitnitsky distribution amplitude
and the logarithmic corrections above the factorization scale
correspond to the lower (upper) solid curves in the spacelike (timelike)
regions.  Similarly, the lower (upper) dashed curves correspond
to our prediction in the spacelike (timelike)
regions when using the asymptotic distribution amplitude.
The parameter $\mu$ was
fixed from the known value of the normalization of the $T$ form factor at
$Q^2=0$.  This is determined from the observed decay width $\pi\to 2\gamma$,
$T(0) \sim 0.14 \mbox{GeV}^{-1}$, which gives
$\mu\simeq 338$~MeV.  This is certainly a mass scale typical
of both confinement and dynamical chiral symmetry breaking as anticipated.
The dotted curve is a simple dipole fit to the data using
the vector meson dominance model (VMD), where
$T(Q^2)=T(0)/[1+(Q^2/M^2)]$
with the parameter choice $M\simeq750$~MeV.
Again, at very large $|Q^2|$ solid and dashed curves converge separately.
In contrast to the pion form factor case, however, this convergence seems
to be much faster, and as shown Figs.~\ref{rgpg}a) and b) the
relative $O(1/Q^4)$ correction to the pQCD predictions,

\begin{equation}
\delta T(Q^2) \equiv { {T_{\rm pQCD}(Q^2) - T(Q^2) }\over
{ T_{\rm pQCD}(Q^2) }}\;,
\end{equation}
is only about $30\%$ ($10\%$) and $20\%$ ($10\%$) for the
Chernyak and Zhitnitsky distribution amplitude and asymptotic one
in a spacelike (timelike) regions respectively.
It is worth noting that the difference in the pQCD and our
predictions for the $T(Q^2)$ form factor below $Q^2 \lsim 50\mbox{GeV}^2$
has opposite sign in the space- and timelike regions. This is due to the
resonance structure in the timelike region which enhances the soft
contributions and in particular introduces an imaginary
part of the form factor. This feature is reproduced in our model
and comes from the complex logarithm in Eq.~(\ref{FF}) for $Q^2 < -\mu^2$.
As $-Q^2$ increases in the timelike region the two solid (dashed) curves
cross and as $-Q^2\to \infty$ they eventually merge as the imaginary part
becomes negligible  [i.e., ${\rm Im}T(Q^2)/{\rm Re}T(Q^2) \to
\log(-Q^2/\mu^2)/-Q^2$] .
\section{Conclusions}
\label{concl}
We have constructed a covariant, gauge-invariant $q\bar q$ amplitude
for the pion in Minkowski space using a perturbative integral
representation.  We have shown that confinement can be implemented
in a straightforward way.
We have also shown that this wavefunction can be readily connected to
the usual pQCD treatment in the asymptotic regime.
The main conclusions of our pion electromagnetic
form factor analysis come from an examination of the
curves in Figs.~\ref{pionff-fig} and ~\ref{rff}.
Comparing our predictions with the pQCD ones
we see that the influence of the soft pion wavefunction
on the hard gluon exchange diagrams can produce large effects.
For example, our calculation of the soft contributions is as large as
$\sim 90\%$ of the purge pQCD result
for $Q^2 \sim 5 \mbox{GeV}^2$ when the CZ distribution is used. 
For the asymptotic distribution our calculation of
the soft contribution is $\sim 50\%$.
The incorporation of nonperturbative effects into a hard quark scattering 
calculation may be somewhat model dependent, but it seems clear that
the effects {\it can} be very significant at intermediate momentum scales. 
Previous treatments of this subject \cite{ISLS}, 
dealt only with the pure pQCD formula of Eq.~(\ref{F-pQCD}) and
in particular they
manifestly break the hard-soft factorization
as the size of the soft corrections is estimated by modifying the hard
scattering amplitude by a soft gluon mass. On the
other hand, in our approach
we study the sensitivity of the form factors to the soft physics
coming from the wavefunctions while keeping the hard amplitude perturbative.
In addition, the present study has been
an attempt to partially address the issues of
Lorentz covariance, gauge invariance, and the nonperturbative
structure of the covariant
amplitude of Eq.~(\ref{Phi-PT}).  Our results for the pion e.m. form
factor appear to support the conclusion that 
the assumption of dominant pQCD contributions to the pion form factor
at moderate $Q^2$ is not well-founded.

We have also shown that the covariant
pion wavefunction can be directly related to the matrix 
element associated with the $\gamma^*\pi^0\to\gamma$ form factor. We have 
analyzed the extrapolation of this form factor away from the asymptotic
region using the nonperturbative covariant pion wavefunction.
Once we fit $T(0)$, using the single nonperturbative scale parameter
$\mu$, we find that $T(Q^2)$ displays a much weaker dependence on the
detailed form
of the pion wavefunction even at low to moderate $Q^2$, (i.e., $|Q^2|\lsim
10$~GeV$^2$).  In particular, we found that the relative,
wavefunction-dependent,
soft corrections to $T(Q^2)$ are less than $\sim 25\%$ for
$|Q^2| \simeq 10$ GeV$^2$ and become even smaller as $|Q^2|$ increases.
Because of that and because the predictions for the form factor
for different types of distribution amplitudes lead to
different predictions, a precision measurement of the
forward angle cross section for the $e+p \to e+p+\gamma$ transition
would be of considerable interest, (e.g. at CEBAF).
In particular, it may help resolve the question of
large peaks predicted by Chernyak and Zhitnitsky for
$x \to \pm 1$.

The apparently conflicting conclusions drawn from the results for the pion e.m.
form factor $F(Q^2)$ and the $\gamma^* + \pi^0 \to \gamma$ from factor
$T(Q^2)$ require cautious interpretation.  The lesson to be learned
from these simple model analyses may be that leading pQCD
treatments can be relatively successful in describing exclusive
amplitudes in the intermediate momentum regime based on the operator product
expansion [$T(Q^2)$], but fail quite badly when there is no appropriate
short distance
expansion involved [e.g., $F(Q^2)$]~\cite{ISLS}.  Further studies along these
lines will be needed before more definitive answers are possible.
\acknowledgements
We would like to acknowledge helpful discussions with D.~Robson,
J.~Piek\-a\-r\-e\-wicz, and A.~Radyushkin.  
This research was supported in part by the U.S. Department of Energy
through Contract Nos. DE-FG05-86ER40273, DE-FG05-86ER40461, DE-FG05-86ER40589,
the Florida State University
Supercomputer Computations Research Institute which is partially funded by
the Department of Energy through Contract No. DE-FC05-85ER250000,
and by the Australian Research Council.
%
%
%

%
%
%
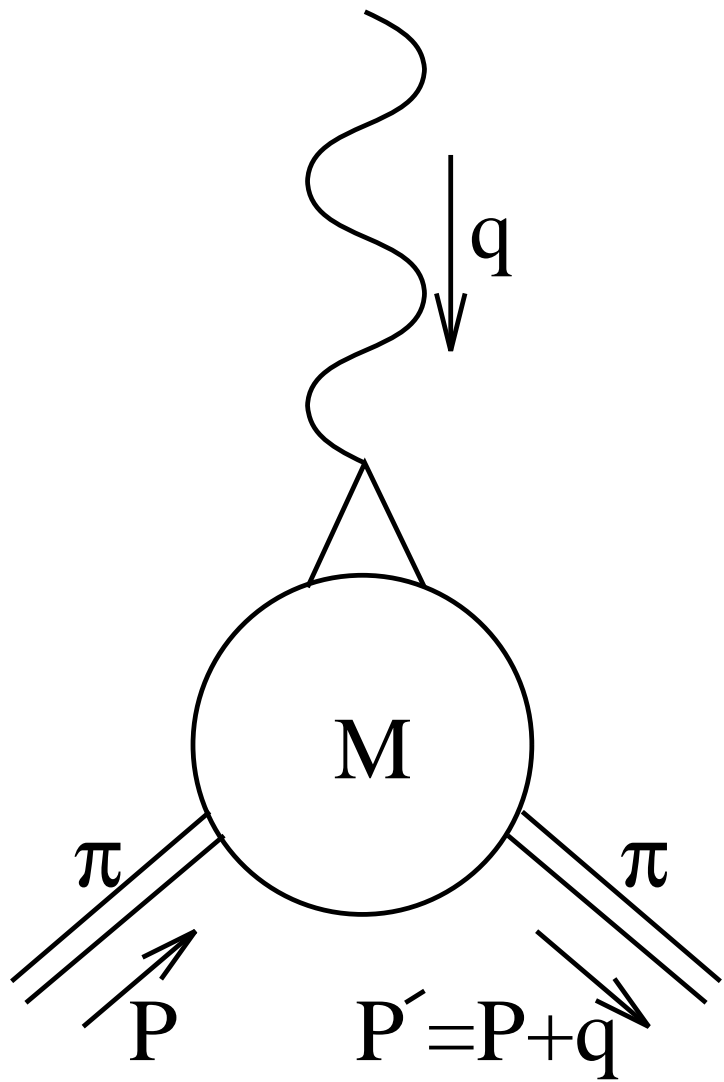
\begin{figure}
\caption{An exact representation of the amplitude for the pion
e.m. form factor.}
\label{pionampl-fig}
\end{figure}
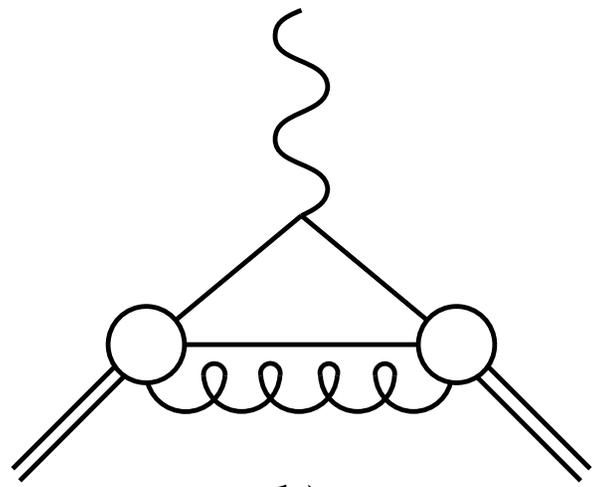
\begin{figure}
\caption{Explicit one-gluon contributions to the high-\protect$Q^2$ region
of the form factor. The pQCD expression of Eq.~(\protect\ref{fact})
is obtained by
multiplying the wave function by the collinear projectors.
a) Hard gluon exchange at the soft pion vertices,
b) spectator gluon (an example spectator diagram).}
\label{oge-fig}
\end{figure}
\begin{figure}
\caption{Spacelike region of the pion e.m .form factor.
The various theoretical curves are explained in the text.
The data are taken from Ref.\protect{\cite{data}}}
\label{pionff-fig}
\end{figure}
\begin{figure}
\caption{Relative size of the nonperturbative contributions to the
pion e.m. form factor for the CZ (solid) and asymptotic (dashed)
distributions.}
\label{rff}
\end{figure}

\begin{figure}
\caption{Diagrams contributing to the large-$|Q^2|$ behaviour of the
$T(Q^2)$ form factor. The pQCD prediction is obtained by multiplying the
wavefunction by a collinear projection. }
\label{gpgpqcd}
\end{figure}

\begin{figure}
\caption{The $\gamma^*\pi \to \gamma$ transition form factor
for a) spacelike and  b) timelike virtual photon momenta.
The various theoretical curves are explained in the text.
The data are also shown~\protect\cite{BEETAL}.}
\label{g-pi-g-fig}
\end{figure}

\begin{figure}
\caption{An illustration of the relative magnitude of
the soft contributions for the $\gamma^*\pi \to \gamma$
transition form factor in the
a) spacelike and b) timelike regions. Again the solid and dashed curves
are for the CZ and asymptotic distributions respectively. }
\label{rgpg}
\end{figure}

%
%
%

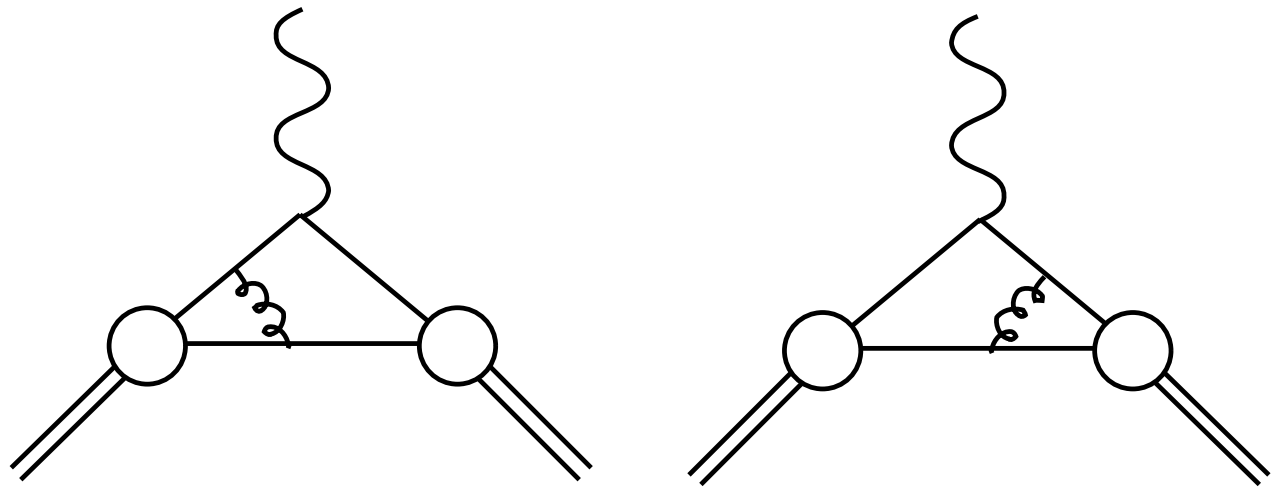
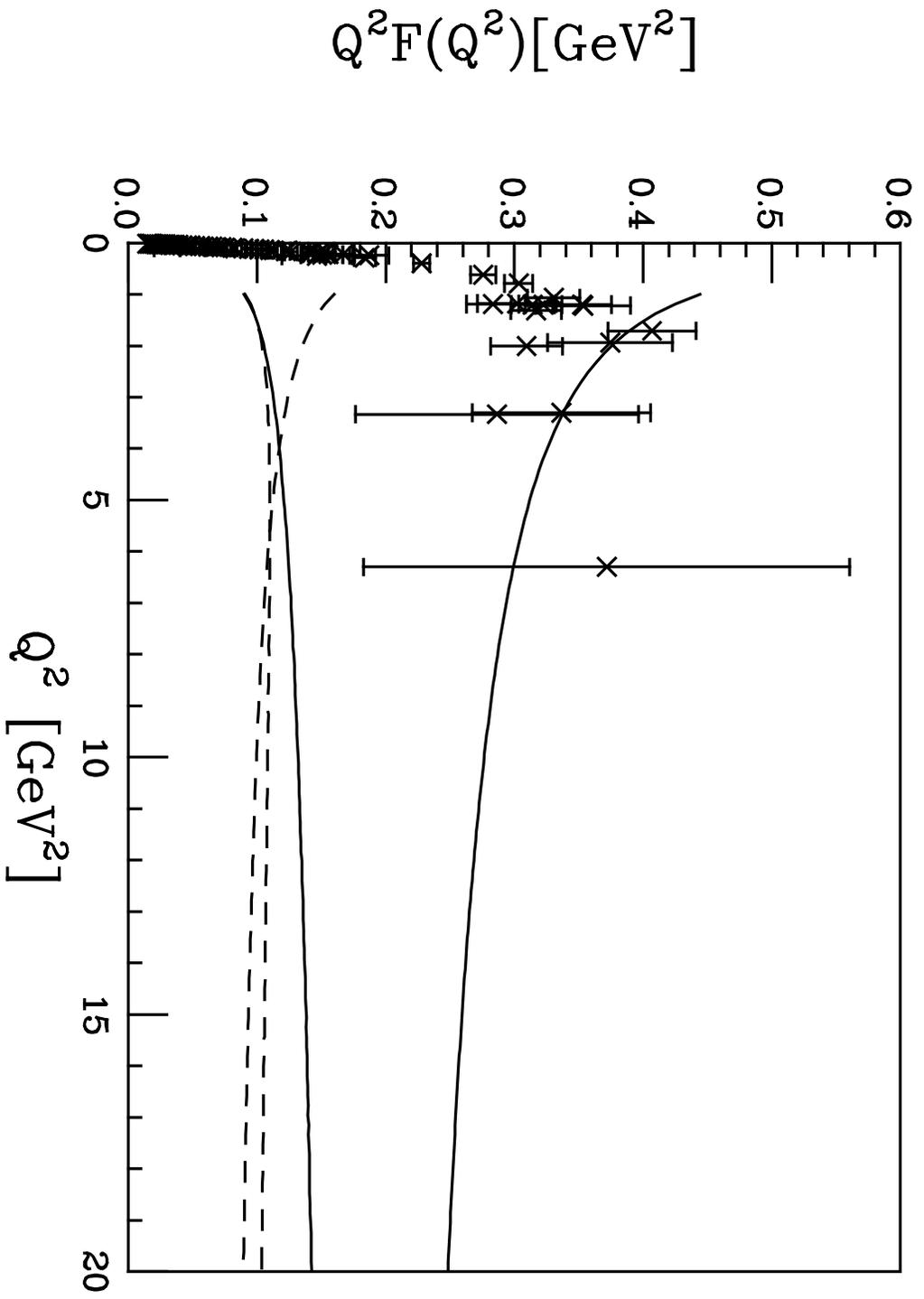
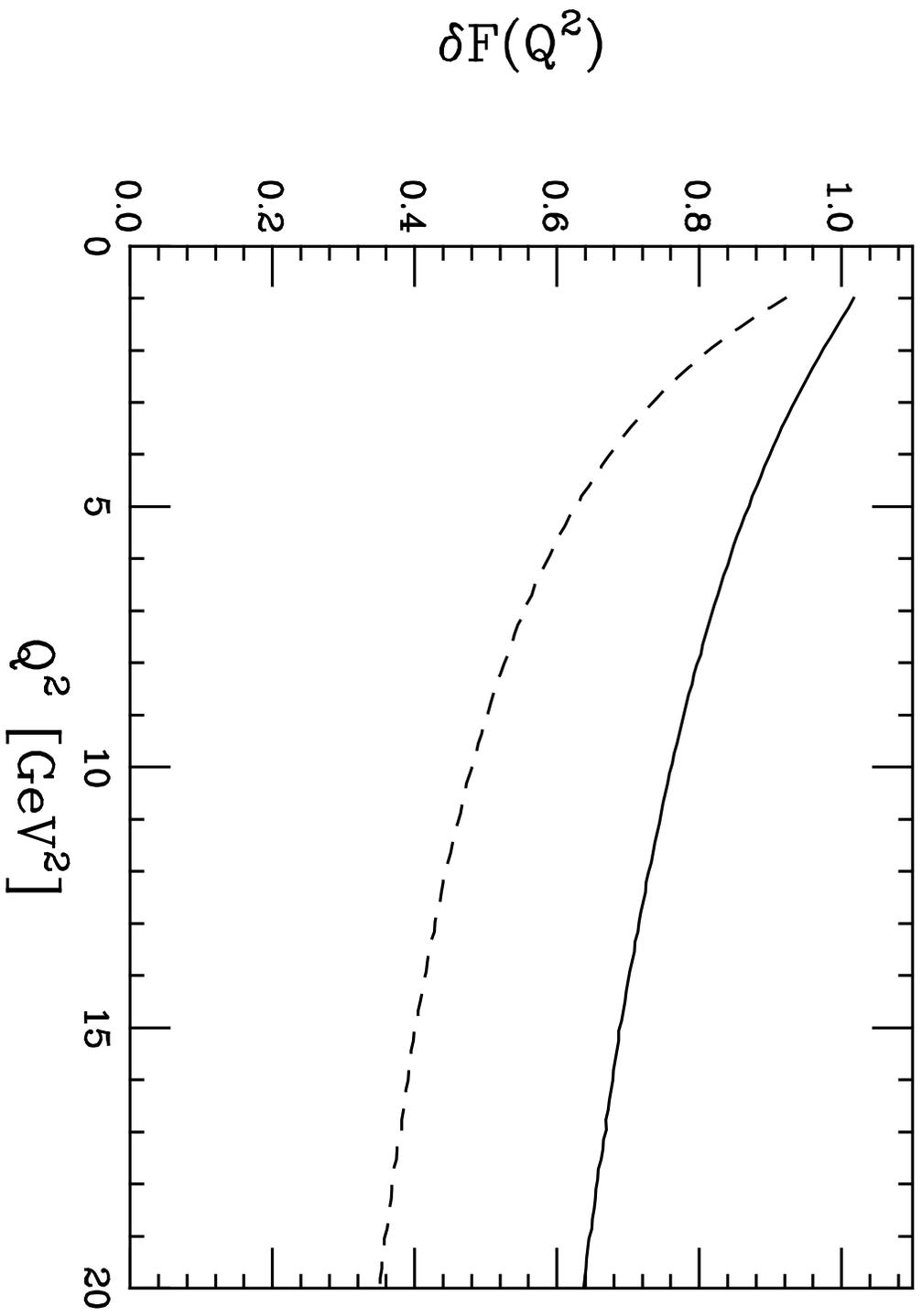
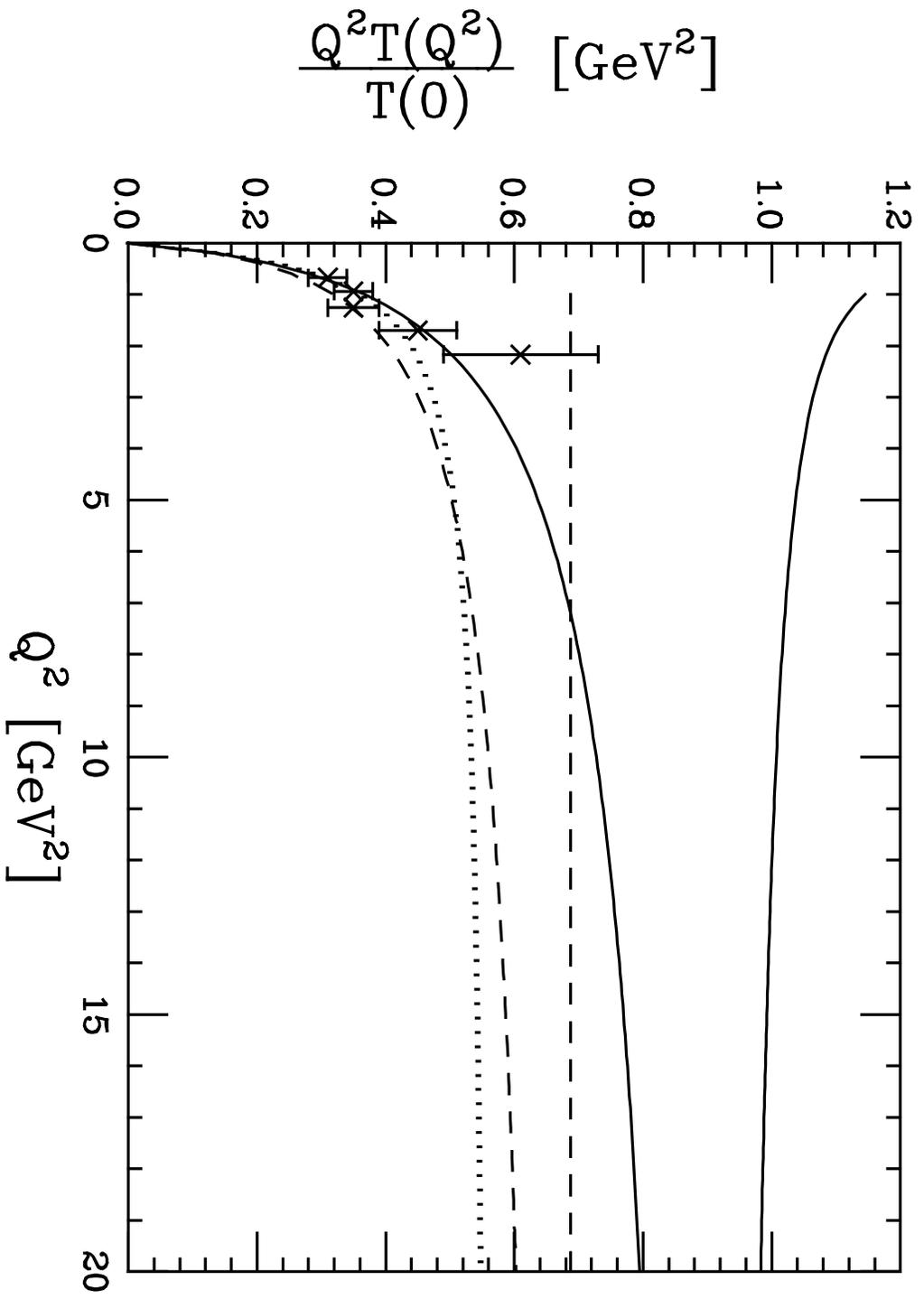
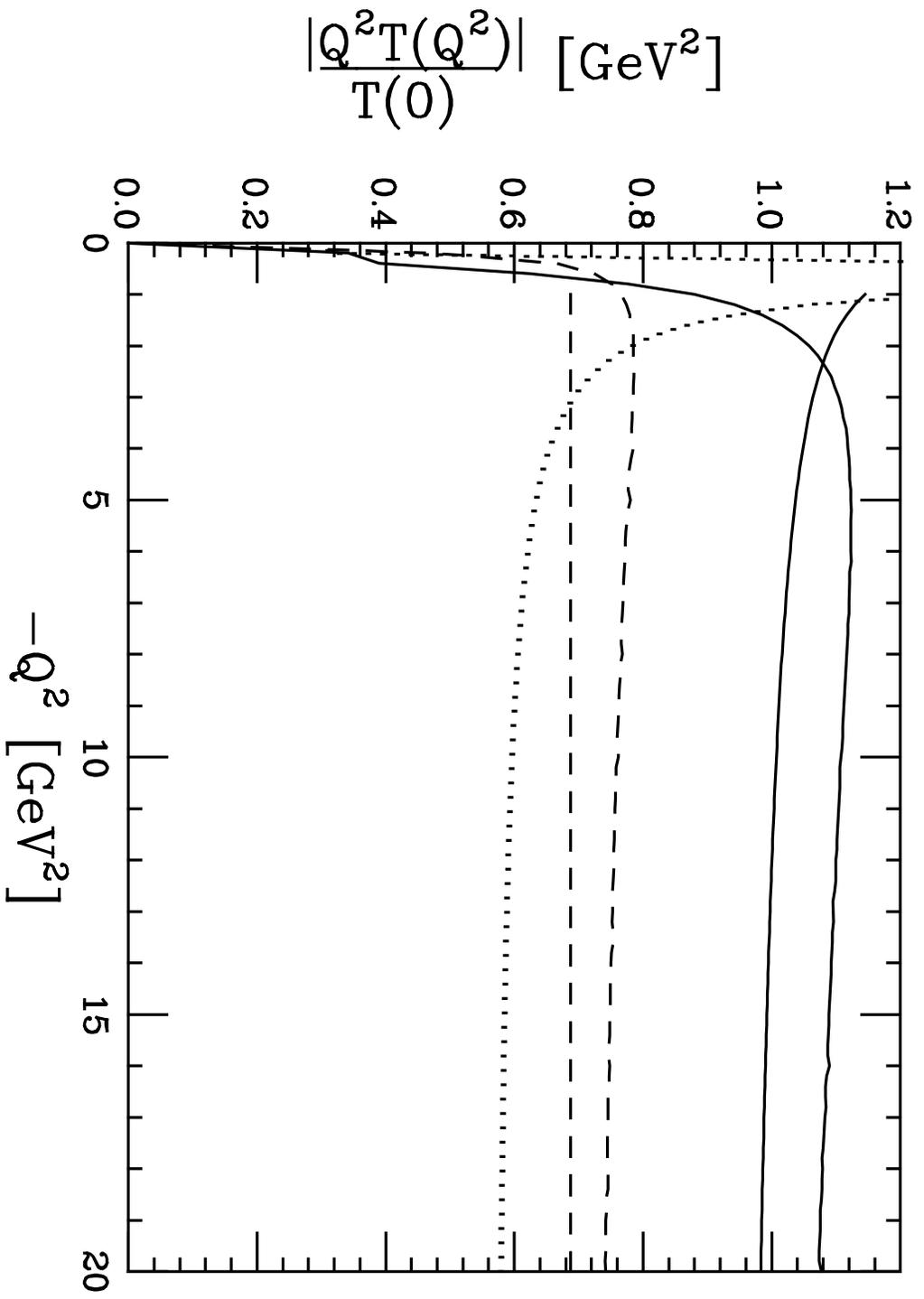
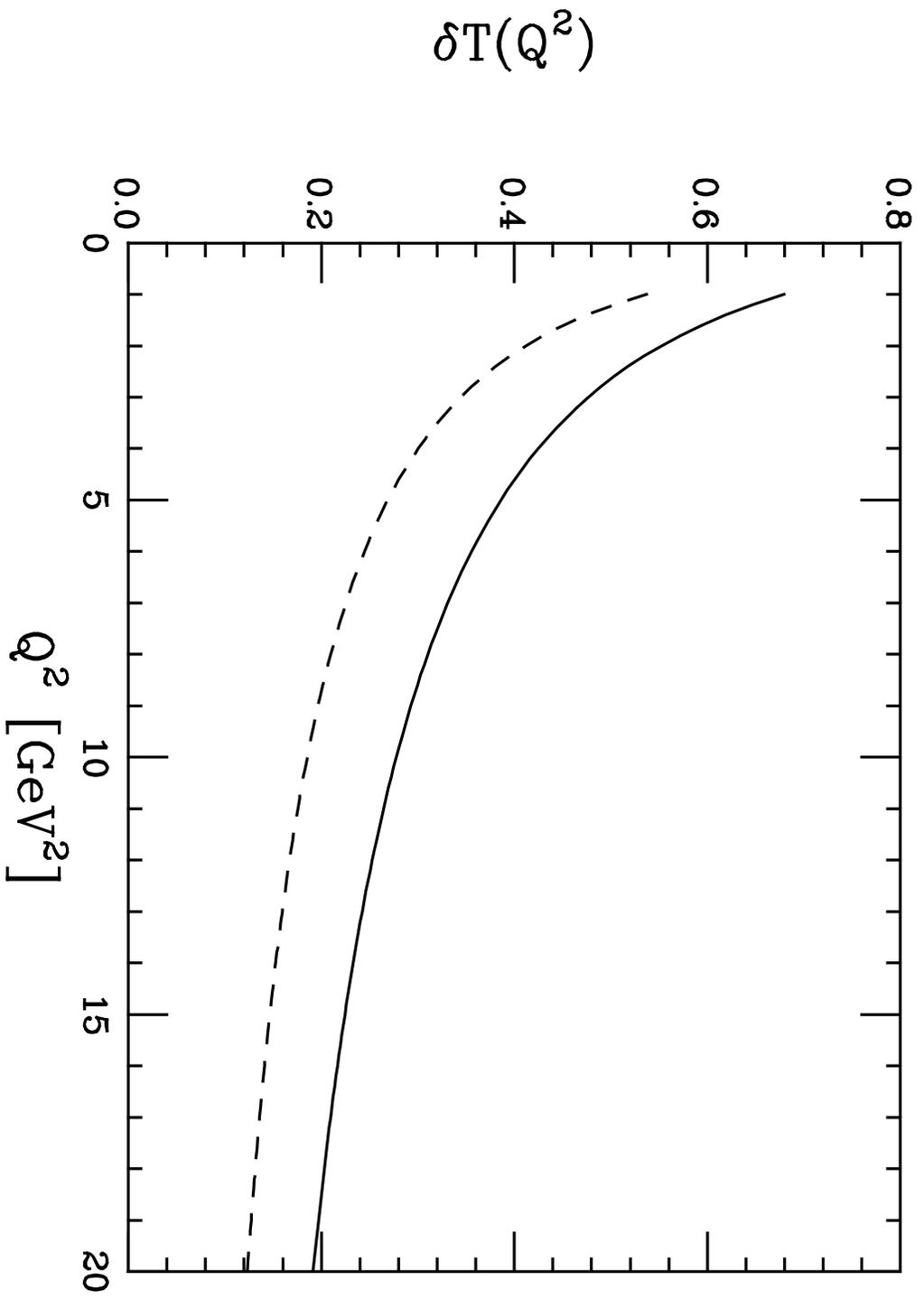
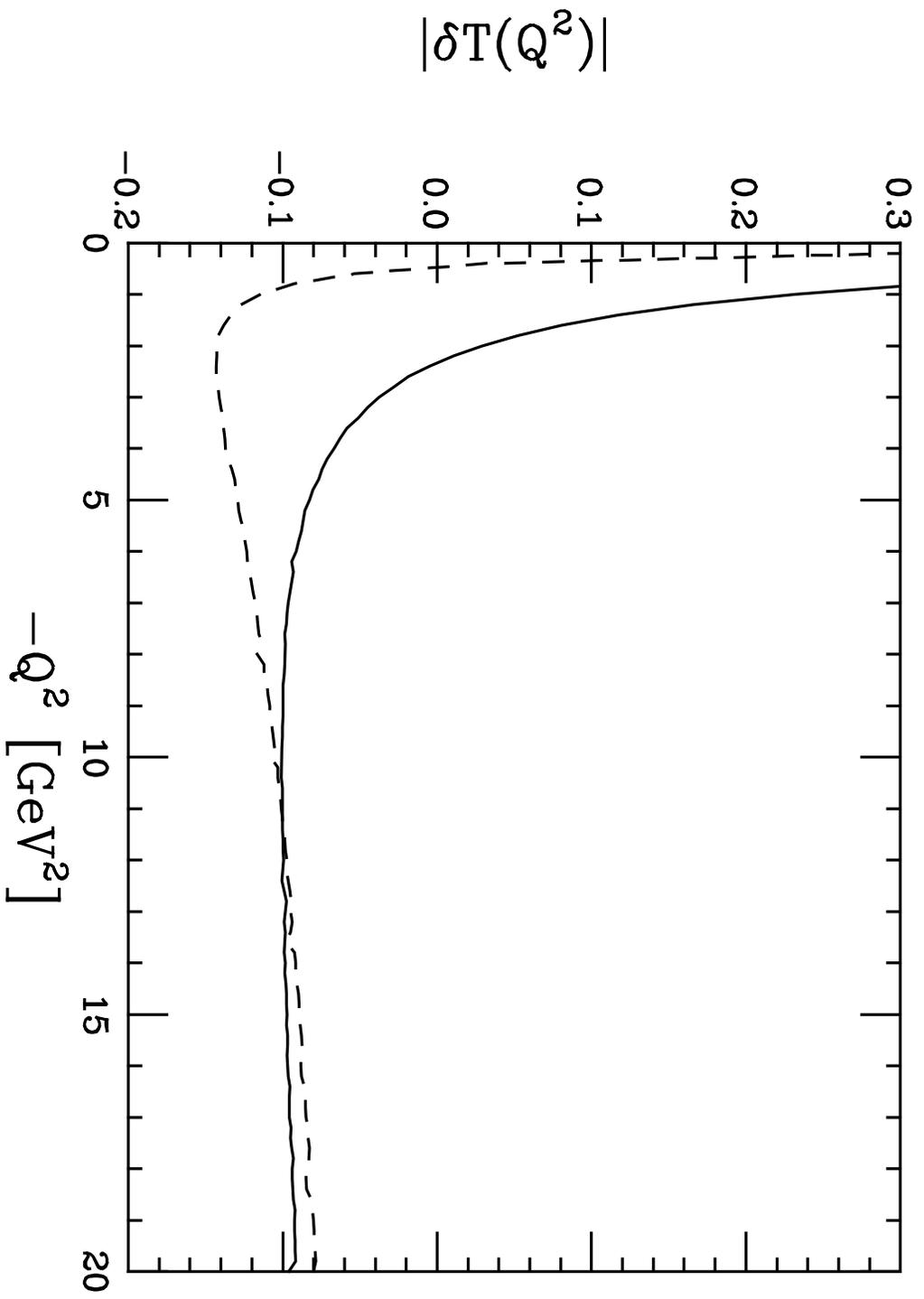
\end{document}